\documentclass[preprint,showpacs,preprintnumbers,amsmath,amssymb]{revtex4}

\usepackage{graphicx,amsfonts}% Include figure files
\usepackage{epsfig}
\usepackage{dcolumn}% Align table columns on decimal point
\usepackage{bm}% bold math
\begin{document}

%\preprint{IFT-P.xx/2009}
%\preprint{ArXiv:yymm.nnnn}
\title{On fermion masses and mixing in a model with $A_4$ symmetry}

% \altaffiliation[Also at ]{Physics Department, XYZ University.}%
\author{A. C. B. Machado}%
\email{ana@ift.unesp.br}
\affiliation{
Instituto  de F\'\i sica Te\'orica--Universidade Estadual Paulista \\
R. Dr. Bento Teobaldo Ferraz 271, Barra Funda\\ S\~ao Paulo - SP, 01140-070,
Brazil
}
\author{J. C. Montero}%
\email{montero@ift.unesp.br}
\affiliation{
Instituto  de F\'\i sica Te\'orica--Universidade Estadual Paulista \\
R. Dr. Bento Teobaldo Ferraz 271, Barra Funda\\ S\~ao Paulo - SP, 01140-070,
Brazil
}
\author{V. Pleitez}%
\email{vicente@ift.unesp.br}
\affiliation{
Instituto  de F\'\i sica Te\'orica--Universidade Estadual Paulista \\
R. Dr. Bento Teobaldo Ferraz 271, Barra Funda\\ S\~ao Paulo - SP, 01140-070,
Brazil
}
\date{16/02/2012
%\today
}% It is always \today, today,
\begin{abstract}
In a recently proposed multi-Higgs extension of the standard model in which discrete symmetries, $A_4$ and $Z_3$ are
imposed we show that, after accommodating the fermion masses and the mixing matrices in the charged currents, the mixing
matrices in the neutral currents induced by neutral scalars are numerically obtained. However, the flavor changing
neutral currents are under control mainly by  mixing and/or mass suppressions in the neutral scalar sector.
\end{abstract}


\maketitle

\section{Introduction}
\label{sec:intro}

One of the main motivations to go beyond the standard model is to have some hints
about the flavor problem, that is, to understand the pattern of fermion masses and mixing.
Most electroweak models have mass matrices of the form
$M_{\alpha\beta}=\sum_i(\Gamma_i)_{\alpha\beta} \langle
\Phi^0_i\rangle$, where the $\Gamma_i$s are, for Dirac fermions, arbitrary complex
dimensionless $3\times 3$ matrices, and $\langle\Phi^0_i\rangle$ denotes the vacuum
expectation values (VEVs) of the neutral scalar fields in the model. For Majorana fermions,
the $\Gamma_i$s are complex symmetric matrices. The mixing matrix and the mass pattern in each
charge sector depend on the structure of the respective $\Gamma_i$s. It is well known that
explicit and predictive forms of these matrices can be obtained by imposing flavor
symmetries.  In particular, most of the \textsl{ansatze} for the fermion mass matrices are of the
form~\cite{xing}
\begin{eqnarray}
M_Q= \left(
\begin{array}{ccc}
E_Q & D_Q& \mathbf{0} \\
D^*_Q  & C_Q & B_Q\\
\mathbf{0}& B^*_Q & A_Q\end{array}\right),
\quad
M^\prime_Q = \left(
\begin{array}{ccc}
\mathbf{0} & D_Q& \mathbf{0} \\
D^*_Q  & C_Q & B_Q\\
\mathbf{0}& B^*_Q & A_Q\end{array}\right),
\label{massmatrices1}
\end{eqnarray}
and other textures which have in common the fact that the zeros appear in symmetric entries.
Recently, it was proposed an extension of the electroweak  model with $A_4$ and $Z_3$ discrete symmetries,
in which the quark mass matrices have the following texture~\cite{dim5}
\begin{eqnarray}
M_Q =\left(
\begin{array}{ccc}
A_Q & \mathbf{0}& D_Q \\
 E_Q & B_Q & \mathbf{0}\\
\mathbf{0}& F_Q& C_Q
\end{array}\right),
\label{usgeral}
\end{eqnarray}
where the zeros occur in non-symmetrical entries.

\section{Quark masses and mixing in model with $A_4$ symmetry }
\label{sec:model}

We consider a model with $G_{SM}\otimes A_4 \otimes Z_3 \otimes Z^{\prime}_3 \otimes
Z^{\prime\prime}_3$ symmetry, where $G_{SM}$ is the standard model (SM) gauge symmetry, $SU(3)_C\otimes SU(2)_L\otimes
U(1)_Y$. In \cite{dim5} quarks and leptons were considered briefly. Here we show a more detailed study in both sectors. Let us begin by considering quarks. The model needs twelve $SU(2)$ doublets,
four triplets of the $A_4$ symmetry: $H\equiv (H_1,H_2,H_3)$, and
$\Phi \equiv (\Phi_1,\Phi_2,\Phi_3)$ for the masses of the $u$-quarks;
$H^\prime\equiv (H^\prime_1,H^\prime_2,
H^\prime_3)$ and
$\Phi^\prime \equiv (\Phi^\prime_1,\Phi^\prime_2,
\Phi^\prime_3) $ for the masses of the $d$-quarks. We also introduce a complex scalar singlet $\zeta$~\cite{dim5}.

Fermion weak eigenstates transform under
$(A_4,Z_3,Z^\prime_3,Z^{\prime\prime}_3)$ as:
\begin{eqnarray}
&&Q_{L} \equiv (Q_{1L},Q_{2L},Q_{3L}) \sim (\textbf{3}, \omega^2, 1, 1);\;\;\quad
 \mathcal{U}_R\equiv (u_{2R},u_{3R},u_{1R}) \sim (\textbf{3}, 1, 1, \omega); \nonumber \\ &&
\mathcal{D}_R\equiv (d_{2R},d_{3R},d_{1R}) \sim (\textbf{3},1, \omega, \omega),
\label{fermions}
\end{eqnarray}
in which $\omega=e^{2\pi i/3}$, and $Q_{iL}=(d_i\,u_i)^T_L,\;i=1,2,3$.  All fermion fields in (\ref{fermions}) are symmetry eigenstates.
Notice that, assuming the usual ordering of $Q_{iL}$, the $A_4$ symmetry will allow, after its breaking, to distinguish among the
right-handed quark components. The scalar fields transform under the same symmetries as:

\begin{eqnarray}
&&
H\sim(\textbf{3}, \omega, 1,  \omega), \;\;
H^\prime\sim(\textbf{3}, \omega^{2}, \omega^{2}, \omega^2),\;
\Phi \sim (\textbf{3},\omega, \omega,
\omega),  \nonumber \\ &&
\Phi^\prime \sim (\textbf{3}, \omega^{2}, 1,  \omega^2),\;
  \zeta \sim (\textbf{1}, 1, \omega,1).
 \label{scalars}
\end{eqnarray}
The notation in the scalar sector is slightly different from that in \cite{dim5}.

With these fermion and scalar fields, and the discrete symmetries above, we have the leading
Yukawa interactions in the quark sector
\begin{eqnarray}
-\mathcal{L}_{Q}= \left[\left(h_{_U}
[\overline{Q}_L\tilde{H}]_A+h^\prime_{_U}\frac{\zeta}{\Lambda}\,
[\overline{Q}_L  \tilde{\Phi}]_B\right)\mathcal{U}_R\right]_1
+\left[\left(h_{_D}[\overline{Q}_L  H^\prime]_A+
h^{\prime}_D \frac{\zeta}{\Lambda} \,
[\overline{Q}_L  \Phi^\prime]_B\right) \mathcal{D}_R\right]_1\!\!+\!\!H.c.
\label{yuka1}
\end{eqnarray}
where $\tilde{\phi}=\varepsilon\phi^*$, with $\varepsilon=i\sigma_2$ ($\sigma_2$ being the usual
Pauli matrix); $[XY]_{A,B}$ means the appropriate product, and $[XY]_1$
denotes the singlet state, see Ref.~\cite{ishimori} and references therein. Instead of the $A$ and $B$ triplet
representations we can use the symmetric, $s$, and antisymmetric, $a$, ones. In fact, $A=s+a$ and $B=s-a$.
We choose to have some fields which products are in  $A$ (or $s$), and others in $B$ (or $a$).
We assume this, firstly, because in global symmetries it is not mandatory to
use all the product representation allowed by the symmetry and, secondly, because that choice may be explained by an
underlying still unknown dynamics. Notice that the $A_4$ symmetry
imposes only two dimensionless Yukawa couplings in each charge sector. We assume that $V/\Lambda\ll
\mathcal{O}(1)$, where  $V$ denotes any VEV  of the model and $\Lambda$ is an energy scale characterizing
an unknown physics. Notice  that the model has flavor changing neutral currents (FCNC) since several Higgs
doublets contribute to the masses of a given charge sector~\cite{gw}.

The renormalizable interactions in (\ref{yuka1}) induce diagonal interactions in the weak basis which explicitly read
\begin{eqnarray}
-\mathcal{L}_{Q}& =& h_{_U} [\overline{Q}_{2 L} \tilde{H}_3u_{2R}+
\overline{Q}_{3 L} \tilde{H}_{1}u_{3R} + \overline{Q}_{1 L}
\tilde{H}_2u_{1R}] \nonumber \\&+&
h_{_D}[\overline{Q}_{2 L}H^{\prime}_3 d_{2R}+ \overline{Q}_{3 L} H^{\prime}_1d_{3R}
+ \overline{Q}_{1 L} H^{\prime}_2d_{1R}] +H.c.
\label{yuka2}
\end{eqnarray}
All scalar doublets are of the form $(x^+_i\,x^0_i)^T$. On the other hand, the non-renormalizable interactions (\ref{yuka1}) are written explicitly as
\begin{eqnarray}
-\mathcal{L}^{nr}_{Q}& =& h^\prime_{_U}\frac{v_\zeta}{\Lambda} [\overline{Q}_{3 L} \tilde{\Phi}_2u_{2R}+
\overline{Q}_{1 L} \tilde{\Phi}_{3}u_{3R} + \overline{Q}_{2 L}
\tilde{\Phi}_1u_{1R}] \nonumber \\&+&
h^\prime_{_D}\frac{v_\zeta}{\Lambda}[\overline{Q}_{3 L}\Phi^{\prime}_2 d_{2R}+ \overline{Q}_{1 L} \Phi^{\prime}_3d_{3R}
+ \overline{Q}_{2 L} \Phi^{\prime}_1d_{1R}] +H.c.
\label{yuka3}
\end{eqnarray}

The mass matrices obtained from Eqs.~(\ref{yuka2}) and (\ref{yuka3}) are
\begin{eqnarray}
M_{_U} \approx h_{_U}\,\left(
\begin{array}{ccc}
v_2 & \mathbf{0}& a_{_U} v_{\phi_3} \\
a_{_U} v_{\phi_1} & v_3 & \mathbf{0}\\
\mathbf{0}& a_{_U} v_{\phi_2}& v_1\end{array}\right)
+H.c.,
\label{us}
\end{eqnarray}
for the $u$-type quarks, and
\begin{equation}
M_{_D} \approx h_{_D}\,\left(\begin{array}{ccc} v^{\prime}_2 &
\mathbf{0}& a_{_D} v^\prime_{\phi_3} \\ a_{_D} v^{\prime }_{\phi_1} &
v^{\prime}_3 & \mathbf{0} \\ \mathbf{0}& a_{_D} v^{\prime }_{\phi_2}&
v^{\prime}_1\end{array}\right)
+ H.c.,
\label{ds}
\end{equation}
for the $d$-type quarks. The texture of these matrices are different from those of other multi-Higgs models
like the private Higgs in \cite{porto1}, and we stress that they are
a consequence of the $A_4$ symmetry and, mainly, of the choice of the representation $A$ or $B$.
Above we have defined $a_{_U} = \frac{h^\prime_{_U}}{h_U} \frac{v_{\zeta}}{\Lambda}$ and $a_{_D} = \frac{h^\prime_{_D}}{h_{_D}} \frac{v_{\zeta}}{\Lambda}$,  for the
$2/3$ and $-1/3$ charged quarks, respectively, and we have denoted $\langle h^{0}_i \rangle=v_i$,
$\langle\phi^{0}_i\rangle=v_{\phi_i}$, $\langle h^{\prime0}_i\rangle=v^{\prime}_i$, $\langle\phi^{\prime0}_{i}
\rangle=v^{\prime}_{\phi_i}$, and $\langle\zeta \rangle=v_\zeta$. For the sake of simplicity, all parameters in
Eqs.~(\ref{us}) and (\ref{ds}) have been considered real. Notice that since each charged sector has its
private VEVs, these matrices are independent from each other.

Here we will show, numerically, that with the mass
matrices (\ref{us}) and (\ref{ds}) it is possible to accommodate the observed masses and the mixing matrices in
the quark sector.  These mass matrices  are diagonalized by bi-unitary transformations,
$V^{U}_LM_{_U}V^{U\dagger}_R=\hat{M}_{_U}$ and $V^{D}_LM_{_D}V^{D\dagger}_R=\hat{M}_{_D}$, respectively, with
$\hat{M}_{_U}=diag(m_u,m_c,m_t)$ and $\hat{M}_{_D}=diag(m_d,m_s,m_b)$. The change of basis is $q_{iL(R)}=
(V^Q_{L(R)})_{i\alpha}q_{\alpha L(R)}$, $q_\alpha$ denotes the quark mass eigenstates
of the respective charge sector, $q_\alpha=u,c,t$ for quarks with electric charge $2/3$ and $q_\alpha=d,s,b$ for
for quarks with electric charge $-1/3$.
We obtain the unitary $V^{U,D}_{L,R}$ matrices, by solving the
matrix equations:
\begin{eqnarray}
&&V^{D}_LM_{_D}M^\dagger_{_D}V^{D \dagger}_L=diag(m^2_d,m^2_s,m^2_b)=(\hat{M}_{_D})^2,\;\;
V^{D}_RM^\dagger_{_D}M_{_D}V^{D\dagger}_R=(\hat{M}_{_D})^2,
\label{vd}\nonumber \\&&
V^{U}_LM_{_U}M^\dagger_{_U}V^{U\dagger}_L=diag(m^2_u,m^2_c,m^2_t)=(\hat{M}_{_U})^2,\;\;
V^{U}_RM^\dagger_{_U}M_{_U}V^{U \dagger}_R=(\hat{M}_{_U})^2,\nonumber \\&&
V^{Q\dagger}_LV^Q_L=\textbf{1},\;\;V^{Q\dagger}_RV^Q_R=\textbf{1},\quad Q=U,D,
\label{vdvu2}
\end{eqnarray}
using as input parameters those VEVs, $a_q$ and $h_q$, appearing in $M_{_U},M_{_D}$,
which give the observed quark masses at an appropriate energy scale (the $Z$ mass in the present case) and the Cabibbo–-Kobayashi–-Maskawa quark--mixing matrix, defined as $V_{_{CKM}}=V^U_LV^{D\dagger}_L$. The latter one is such that, at least in the context of the SM, the mixing between the first two families is almost scale independent and for the third family $V_{cb}$ and $V_{ub}$ change at the level of 13-16\% between $m_t$ and $10^{15}$ GeV~\cite{ckmrg}. Here, therefore the $V_{_{CKM}}$  will be considered scale independent.

Since the mass matrices (\ref{us}) and (\ref{ds}) are predictions of the model, they are valid at the energies at which all symmetries of the model are realized, \textit{i.e.}, at the electroweak scale. For this reason, we use the running quark masses at $\mu=M_Z$, taken from Ref.~\cite{massas}, for light quarks (in MeV): $m_u=1.27^{+0.50}_{-0.42}$, $m_d=
2.90^{+1.24}_{-1.19}$, $m_s=55^{+16}_{-15}$; and for heavy quarks (in GeV): $m_c=0.619\pm 0.084$,  $m_b=2.89\pm 0.09$,
$m_t=171.7\pm3.0$.

In order to obtain the values of the $V^{U,D}_{L,R}$ matrix elements within an interval,
we find two sets of values of the input parameters which give the quark masses and the CKM entries within
the experimental errors. For instance, using:  1) $h_{_D} = 0.1$ and $a_{_D}=0.11$
and (all VEVs are given in GeV) $v^{\prime}_1=28.9$, $v^{\prime}_2= 0.44$, $v^{\prime}_3=0.03$,
$v^{\prime}_{\phi_1}=1.95$, $v^{\prime}_{\phi_2}=0.03$, and $v^{\prime}_{\phi_3}= 8.1$,  and 2) $h_{_D} =
0.1$ and $a_{_D}=0.2$ and $v^{\prime}_1=29.8$, $v^{\prime}_2=0.54$, $v^{\prime}_3=0.03$,
$v^{\prime}_{\phi_1}=0.35$, $v^{\prime}_{\phi_2}=0.01$, and $v^{\prime}_{\phi_3}= 2.9$, we obtain the following
values for the masses in the $d$-quark sector: $m_d= (2.70-2.97)$ MeV, $m_s=(48.95-54.4)$ MeV, and $m_b=
(2.89- 2.98)$ GeV. For the $u$-quark sector, we use 1) $h_{_U} =1.11$, $a_{_U}=0.2$, and the VEVs $v_1=153,
\,v_2=0.54,\,v_3=0.001125$, $v_{\phi_1}=0.08875,\,v_{\phi_2}=0.03555,\,v_{\phi_3}=57.2355$; 2) $h_{_U} =1.11$,
$a_{_U}=0.13$, and the VEVs $v_1=153,\,v_2=0.531,\,v_3=0.00108$, $v_{\phi_1}=1.2048,\,v_{\phi_2}=
0.3199,\,v_{\phi_3}=70.783$, we obtain: $m_u=(1.27 - 1.93)$~MeV, $m_c=(598- 613)$ GeV, and
$m_t=(170.305- 170.137)$ GeV. In spite that some VEVs are small, this does not imply necessarily the existence of
light scalars. See Ref.~\cite{3Dmodel} where the case of three Higgs scalar doublets with $A_4$ symmetry were considered in details.

The matrices $V^{U,D}_{L,R}$ can be obtained by solving (\ref{vdvu2}) with
values for the parameters considered above.
An example of numerical $V^{U,D}_L$ matrices obtained is given by:
\begin{eqnarray}
&&
V^U_L = \left(
\begin{array}{ccc}
0.03290 \to 0.28285 & - (0.95902 \to 0.99946) & - (0.00246 \to 0.01678)\\
- (0.95733 \to 0.99667)& - (0.0330 \to 0.28334) & 0.05680 \to 0.07457 \\
0.05923\to 0.07461 & (0.3 \to 2)\times 10^{-7}& 0.99721\to 0.99824 \end{array}\right),
\nonumber \\&&
V^D_L = \left(
\begin{array}{ccc}
0.12895\to 0.25492 &-( 0.96693\to 0.99165) & -(0.00251\to 0.00786)\\
-(0.96647\to0.99146)& -(0.12896\to 0.25504) & 0.0193\to0.0298 \\
0.01946\to 0.03082& (2\to9) \times 10^{-7}& 0.99952\to0.99981\end{array}\right).
\label{vudl}
\end{eqnarray}
Using these matrices and the definition $V_{_{CKM}}=V^U_LV^{D\dagger}_L$, we obtain
\begin{equation}
\vert V_{_{CKM}}\vert = \left(
\begin{array}{ccc}
0.9748- 0.9875  & 0.1571- 0.2230  & 0.0014 - 0.0113 \\
0.1574 - 0.2227 & 0.9739 - 0.9868  & 0.0381- 0.0438 \\
0.0051- 0.0111 & 0.0394 - 0.0424& 0.9990- 0.9992\end{array}\right),
\label{ckm}
\end{equation}
We compare this matrix with the global fit of the magnitude of the $V_{_{CKM}}$ elements given in Eq.~(11.27)
of PDG~\cite{pdg}
\begin{equation}
\vert V^{pdg}_{_{CKM}}\vert =\left(\begin{array}{ccc}
0.97428\pm0.00015\, & 0.2253\pm0.0007& 0.00347^{+0.00016}_{-0.00012}\\
0.2252\pm0.0007&\, 0.97345^{+0.00015}_{-0.00016} &0.010^{+0.0011}_{-0.0007}\\
0.00862^{+0.00026}_{-0.00020}\, & 0.0403^{+0.0011}_{-0.0007} & 0.999152^{+0.000030}_{-0.000045}\\
\end{array}\right).
\label{ckmpdg1}
\end{equation}
Considering the elements of $V^{pdg}_{_{CKM}}$,
we note that the $V_{_{CKM}}$ in Eq.~(\ref{ckm}) has all its entries within
1-$\sigma$ but the $V_{td}$, which is within 1.7-$\sigma$.
We recall that at present there are several discrepancies between experiments and
the standard model at the tree level that are about 3$\sigma$ standard deviations~\cite{lenz}.
%%%%%% CP INSERTION 
Notice that we are not considering  CP violation in our analysis. This is because 
in our framework the CP violation issue is more complicated than in the
SM case since in our  model there are more CP violation phases.
Although we can perform the usual phase redefinition in the left-handed fields of the charged currents, in order to obtain the full, single--phase, $V_{CKM}$ matrix, many  CP violating phases will still be present in the Yukawa interaction  Lagrangian. We find that this subject
deserves a separate study.

In the same way we have obtained, from (\ref{vdvu2}), the respective $V^{U,D}_R$ matrices:
\begin{eqnarray}
&& V^U_R = \left(
\begin{array}{ccc}
-(0.00091\to0.00371) & -(0.99999\to 1.) & 0.00005\to0.00027\\
0.99999\to1.& 0.00091\to0.00372 & -(0.0002\to0.00026) \\
0.0002\to 0.00026& 0.00004\to0.00027 & 1.\end{array}\right),\nonumber \\&&
V^D_R = \left(
\begin{array}{ccc}
0.00047\to0.01661 & -(0.99986\to0.99997) & 0.00006\to 0.00011\\
-(0.99986\to0.99997)& -(0.00703\to0.01661)& 0.00035\to0.00047 \\
0.00035\to0.00047 & 0.00007\to0.00011& 1.\end{array}\right).
\label{vudr}
\end{eqnarray}
The matrices above will appear in flavor changing neutral currents in the Yukawa interactions.

Notice that the mass matrices in Eqs.~(\ref{us}) and (\ref{ds}) have the following parameters (assuming all of
them to be real): $h_{_U},a_{_U},h_{_D},a_{_D}$ and twelve VEVs. It means 16 real free parameters
to explain 12 mixing angles of four matrices: $V^{U,D}_L$ (or one of $V^{U,D}_L$ and $V_{_{CKM}}$) and $V^{U,D}_R$.
However, if $CP$ violation is allowed, it is necessary to chose a weak
basis~\cite{weakbasis,waponce} in order to eliminate some extra phases which may be difficult to measure.

%\subsection{FCNC suppression}
%\label{subsec:fcnc}

\textsl{FCNC suppression}. There are flavor changing neutral currents (FCNC) effects in both quark sectors. Here we will consider only some of these sort of effects which occurs at the tree level.
For instance, from (\ref{yuka2}) the renormalizable interactions of the doublets with the $d$-type quarks
(in the weak basis), are given by $h_{_D}(\bar{d}_{2L}s_{2R}\varphi^{\prime0}_3+\bar{d}_{3L}d_{3R}
\varphi^{\prime0}_1+\bar{d}_{1L}d_{1R}\varphi^{\prime0}_2)$. The other three doublets have interactions,
which are suppressed by the scale $\Lambda$, will not be considered.
On the other hand, the neutral scalar weak eigenstates $\varphi^{\prime0}_\alpha$ are
linear combinations of the neutral scalar mass eigenstates, $h^0_n$, i.e., $\varphi^{\prime0}_\alpha=\sum_{n}
U_{\alpha n}h^0_n$, $\alpha,\,n=1,2,3$. This is in fact a simplification, since the model has several doublets,
but it may be assumed (for the sake of simplicity) that in each charge sector
there are three doublets which give the more important effects. It means that the mixing matrices in the full scalar sector may be almost block diagonal, with each one related to a given fermion charge sector.

The suppression of FCNC can be obtained at least with a reasonable fine tuning in the mixing parameters or by considering heavy enough the charged and neutral scalars. We consider some examples just for illustrating this point.
The contributions of the neutral scalars to the $\Delta M_K$ put the strongest constraint on some of the parameters
of the model. From (\ref{yuka2}), the renormalizable Yukawa interactions between the $d$ and $s$ quarks and a given
neutral (pseudo)scalar, denoted by $h^0_n(A^0_n)$, are given by
\begin{equation}
-\mathcal{L}_{ds}=h_{_D}\sum_n[\bar{d}_LK_ns_R+\bar{s}_LK^\prime_nd_R](h^0_n+iA^0_n)+H.c.
\label{newint1}
\end{equation}
where we have assumed $h_{_D}$ real and defined
\begin{eqnarray}
&&K_n=(V^D_L)^*_{sd}(V^D_R)_{ss}U_{3n}+(V^D_L)^*_{bd}(V^D_R)_{bs}U_{1n}+(V^D_L)^*_{dd}(V^D_R)_{ds}U_{2n}\nonumber \\&&
\approx -0.01U_{3n}+ 0.16U_{2n}+10^{-9}U_{1n}, \nonumber \\&&
K^\prime_n=(V^D_L)^*_{ss}(V^D_R)_{sd}U_{3n}+(V^D_L)^*_{bs}(V^D_R)_{bd}U_{1n}+(V^D_L)^*_{ds}(V^D_R)_{dd}U_{2n}\nonumber \\&&
\approx 0.01U_{3n}-0.01U_{2n}+10^{-8}U_{1n},
\label{kas}
\end{eqnarray}
where we have used the values of the $V^D_L$ and $V^D_R$ matrix elements given in (\ref{vudl}) and (\ref{vudr}), respectively.

Thus, the interactions in (\ref{newint1}) can be rewritten as
\begin{eqnarray}
&&-\mathcal{L}_{sd}=\frac{h_{_D}}{2}\{[(K_n+K^{\prime*}_n)(\bar{d}s)+(K_n-K^{\prime*}_n)(\bar{d}\gamma_5s)]
(h^0_n+iA^0_n)]\nonumber \\&&
+[(K^*_n+K^\prime_n)(\bar{s}d)-(K^*_n-K^\prime_n)(\bar{s}\gamma_5d)](h^0_n+iA^0_n)^*\},
\label{newint2}
\end{eqnarray}
and the effective Hamiltonian contributing to $K^0\leftrightarrow \bar{K}^0$ transition is given by
\begin{eqnarray}
\mathcal{H}^{\Delta S=2}_{eff}\vert_{scalars}=\sum_n\frac{h^2_{_D}}{4m^2_n}[(K^*_n+K^\prime_n)^2(\bar{s}d)^2+
(K^*_n-K^\prime_n)^2(\bar{s}\gamma_5d)^2]
\label{heff}
\end{eqnarray}

In the vacuum insertion approximation $\langle \bar{K}^0\vert (\bar{s}d)^2_{V-A} \vert K^0\rangle=2M_Kf^2_K/3$.
Using (up to some phases)~\cite{branco,ciuchini}
\begin{eqnarray}
&&\langle \bar{K}^0\vert (\bar{s}d)^2 \vert K^0\rangle=-\frac{f^2_KM_K}{12}\left[1-\frac{M^2_K}{(m_s+m_d)^2} \right],
\nonumber\\&&
\langle \bar{K}^0\vert (\bar{s}\gamma_5d)^2 \vert K^0\rangle=\frac{f^2_KM_K}{12}\left[1-11 \frac{M^2_K}{(m_s+m_d)^2}\right],
\label{elementos}
\end{eqnarray}
we obtain the following extra contributions to $\Delta M_K$ due to the neutral scalar
\begin{eqnarray}
\Delta M_K\vert_{scalars}=2\textrm{Re}\langle \bar{K}^0\vert \mathcal{H}^{\Delta S=2}_{eff}\vert_{scalars}\vert K^0\rangle= \textrm{Re}\sum_n\zeta^n_{sd}\frac{2}{3}
M_Kf^2_K,
\label{deltak}
\end{eqnarray}
where
\begin{eqnarray}
\textrm{Re}\sum_n\zeta^n_{sd}&=&\frac{h^2_D}{8}\sum_n\frac{1}{m^2_n}
\textrm{Re}\left\{-(K^*_n+K^\prime_n)^2
\left[1-\frac{M^2_K}{(m_s+m_d)^2}\right]+(K^*_n-K^\prime_n)^2\left[1-11\frac{M^2_K}{(m_s+m_d)^2}\right]\right\}\nonumber
\\&\approx&-\textrm{Re}\sum_n\frac{ 7.52(U^*_{3n}+U_{3n})U^*_{2n}+ 0.75U^2_{2n}}{(m_n/100\textrm{GeV})^2}\,10^{-5}\,\textrm{GeV}^{-2},
\label{zetan}
\end{eqnarray}
where we have used only the main contributions in (\ref{kas}). There are also similar contributions induced by the pseudoscalar, $A^0_n$. For illustrative purposes we showed above only the scalar contributions.

Notice that, in the present context, in (\ref{kas}) only the matrix elements $U_{\alpha n}$ are not known yet and that, independently of the mass eigenstates $h^0_n$, $U_{1n}$ will not be constrained by processes like $\Delta M_K$. In order to be consistent with data $\textrm{Re}\,\sum_n\zeta^n_{sd}$ must be smaller than the contribution of the SM: i.e., $\textrm{Re}\,\zeta^{SM}_{sd}= G^2_Fm^2_c\textrm{Re}\,[(V_{CKM})^*_{cd}(V_{CKM})_{cs}]^2/16\pi^2\approx 10^{-14}\,\textrm{GeV}^{-2}$ (we have used only the dominant contribution of the $c$ quark and $g(m_c/M_W)=1$). By imposing that
$\textrm{Re}\,\sum_n\zeta^n_{sd}<\textrm{Re}\,\zeta^{SM}_{sd}$, implies, from (\ref{zetan})
\begin{equation}
\left\vert\frac{15\textrm{Re}\,U_{3n}\cdot \textrm{Re}\,U_{2n}+0.75\left[(\textrm{Re}\,U_{2n})^2-
(\textrm{Im}\,U_{2n})^2\right)]}{(m_n/100\textrm{GeV}]^2}\right\vert
<10^{-9}.
\label{oban}
\end{equation}
Even if for a given $n$, $m_n=100$ GeV, there are two different ways
for each term in (22) to satisfy the constraint:
a) all parameters involved are of the order of $10^{-5}$, i.e. $\textrm{Re}\,U_{3n}
\sim \textrm{Re}\,U_{2n} \sim \textrm{Im}\,U_{2n} \sim 10^{-5}$,  or b) there is a fine tuning
among the parameters. For instance,  $\textrm{Re}\,U_{3n} =0.01, \textrm{Re}\,U_{2n} =0.01,
\textrm{Im}\,U_{2n}=0.045826742$. For heavier neutral scalars the $U_{\alpha n}$ matrix elements may be greater, limited only by the unitarity of the matrix.

Similar constraints come from $\Delta M_B$ and $\Delta M_{B_s}$ data and
other $\Delta B=1$ weak processes, like $B_{s,d}\to\mu^+\mu^-$ and $B^+\to K^+\mu^+\mu^-$ decays.
Experimentally, $\Gamma(B^+~\to~K^+\mu^+\mu^-)/\Gamma_{total}<5.2\times10^{-7}$~\cite{pdg}. In this case,
the quark interactions involved are $h_{_D} \overline{b}_L\left[(V^D_L)^*_{sb} (V^D_R)_{ss}   \varphi^{\prime 0}_{3} +
(V^D_L)^*_{bb} (V^D_R)_{bs} \varphi^{\prime 0}_{1}  +  (V^D_L)^*_{db} (V^D_R)_{ds}
\varphi^{\prime 0}_{2}\right]s_R+H.c.$, and we recall that, in this model, leptons have their own scalar sector
with the Yukawa interactions $g_l [[\overline{L} \hat{H}^\prime]_A l_R]_1$. Here, $\hat{H}$
denotes the triplet of $A_4$ formed by three scalar doublets (see below). For muons it means $g_l \overline{\mu}_L \mu_{R}\hat{H}^0_{3}$, and the muon mass is given by $m_\mu=g_l\langle \hat{h}^0_3\rangle\equiv g_l\hat{v}_3$. In the scalar potential the scalar triplet $\widehat{H}$ mixes with the scalar triplets related to the quark sector only by quartic terms like $\lambda_{1} \left[ \hat{H}^{\dag} \hat{H} \right]_A \left[H^{\prime \dagger} H^\prime \right]_A =
\lambda_{1} \left[ \hat{H}^\dagger_2 \hat{H}_3 H^{\prime \dagger}_2 H^\prime_3 +
\hat{H}^{\dag}_3 \hat{H}_1 H^{\prime \dagger}_3 H^\prime_1 + \hat{H}^{\dagger}_1
\hat{H}_2 H^{\prime \dagger}_1 H^\prime_2    \right]$. This implies terms like $\lambda_1v_2\hat{v}_3$ in the
mass matrix in the neutral scalar sector and, in the weak basis a mass insertion implies that the propagator becomes
$\lambda_1v^\prime_2\hat{v}_3/m^4$, where $m$ denotes a typical
value for the neutral scalar masses. As we said before, in that basis, the Higgs scalars coupled to leptons are different from those coupled to quarks. It means that in semi-leptonic decays a detailed
analysis may be done only if we consider at least six doublets of scalar Higgs bosons. Here, just for illustration consider $b\to s\mu\mu$ decay. Working in the weak basis we obtain,
\begin{equation}
\frac{\Gamma(b\to s\mu^+\mu^-)}{\Gamma(b\to c\bar{\nu}_\mu\mu)}\propto \frac{\lambda^2_1h^2_D\left\vert(V^D_L)^*_{db}
(V^D_R)_{ds}\right\vert^2m^2_\mu v^{\prime\, 2}_2}{G^2_Fm^8}=1.155\times 10^4 \lambda^2_1/(m/\textrm{GeV})^8< 10^{-7},
\label{beleza}
\end{equation}
and we have used the mixing matrix elements $(V^D_L)_{db}$ and
$(V^D_R)_{ds}$ from (\ref{vudl}) and (\ref{vudr}), respectively. All parameters in (\ref{beleza}) but $\lambda_1$ and $m$ are already known. It implies $\lambda^2_1/(m/\textrm{GeV})^8~<~10^{-11}$ which is valid for a wide range of the parameters. For instance, for $m>10$ GeV  and $\lambda^2_1\approx10^{-3}$ the above condition is satisfied.

The case of FCNCs in the $u$-quark sector involves the matrices $V^U_L$ and $V^U_R$. The a\-na\-ly\-sis of the FCNC
in this model is similar to that of Ref.~\cite{ma} in which two doublets were considered.

\textsl{The lepton sector}. In the lepton sector the discrete symmetries are the same as in the
quark sector. Leptons transform under $A_4\otimes(Z_3)^3$ as follows:
\begin{eqnarray}
L\equiv (L_e,L_\mu,L_\tau) \sim (\textbf{3}, \omega, \omega^2, 1);\quad
\;\;l_R\equiv(\mu_R,\tau_R,e_R) \sim (\textbf{3},1, 1, \omega),
\label{lc}
\end{eqnarray}
and the leptophilic scalars transform under those symmetries as
\begin{eqnarray}
&&H^{\prime \prime}\equiv (H^{\prime \prime}_1,H^{\prime \prime}_2,H^{\prime \prime}_3)\sim (\textbf{3},\omega,
\omega^2, 1),\;\; \hat{H}\equiv(\hat{H}_1,\hat{H}_2,\hat{H}_3)\sim (\textbf{3},\omega, \omega^2,  \omega^2),
\nonumber \\ &&
\Phi^{\prime \prime}\equiv (\Phi^{\prime \prime}_1,\Phi^{\prime \prime}_2,\Phi^{\prime \prime}_3)
\sim (\textbf{3}, 1, 1, 1),\;
 \mathcal{T}\equiv (\mathcal{T}_1,\mathcal{T}_2,\mathcal{T}_3)\sim (\textbf{3},
 \omega^{2}, \omega, 1).
 \label{scalars2}
\end{eqnarray}

The Lagrangian of the lepton sector is given by (see the notation in \cite{dim5}):
\begin{eqnarray}
\mathcal{L}&=&\left( g_l [\overline{L} \hat{H}]_A + \frac{g^\prime_l}{\Lambda^2} [\overline{L} \hat{H}]_A
\vert\zeta\vert^2 + \cdots \right)
l_R \nonumber \\ &+&
\frac{1}{\Lambda}\left( f_\nu\,[(\overline{L^c}\epsilon H^{\prime\prime})]_A
[( L\epsilon \Phi^{\prime\prime})]_B
+ \frac{f^\prime_\nu}{\Lambda}
[\overline{L^c} \varepsilon\vec{\sigma}
\cdot \vec{\mathcal{T}}]_B[L\Phi^{\prime\prime\dagger} ]_B\chi+\dots \right)+ H.c.,
\label{lagrangianaleptons}
\end{eqnarray}
where $g_l,g^\prime_l,f_\nu$ and $f^\prime_\nu$ are dimensionless Yukawa couplings and $\Lambda$ an energy scale which may be, or not, the same as that in the quark sector.
The mass matrix for the charged leptons is almost diagonal: $M_l=g_l\textsl{diag}(\hat{v}_2, \hat{v}_3, \hat{v}_1)+
\mathcal{O}(g^\prime_lv^2_\zeta \hat{v}_i/\Lambda^2)$, where $\langle\hat{H}_i\rangle=\hat{v}_i$. In this case the renormalizable
interactions are dominant, i.e., $g_l\hat{v}_1\simeq m_\tau$, $g_l\hat{v}_3~\simeq~m_\mu$ e $g_l\hat{v}_2\simeq~m_e$. Neglecting the
contributions proportional to $g^\prime_l$, the charged lepton mass matrix is diagonal, thus the values of the VEVs are easily
obtained: Just for illustrating, by using the central value for the lepton masses (in MeV) at the $Z$ pole scale:  $m_e\approx 0.486$,
$m_\mu\approx102.718$ and $m_\tau\approx1746.24$, hence $\hat{v}_2\sim m_e/g_l$,  $\hat{v}_3 \sim m_\mu/g_l, \hat{v}_1 \sim m_\tau/g_l$.

On the other hand, the mass matrix for the neutrinos is
\begin{equation}
M_\nu \approx \left(
            \begin{array}{ccc}
             \delta_1&
             \frac{v^{\prime\prime}_2}{v^{\prime\prime}_1} & \frac{v^{\prime\prime}_{\phi_3} }{ v^{\prime\prime}_{\phi_1} } \\
              \frac{v^{\prime\prime}_2}{v^{\prime\prime}_1}  &   \delta_2 &
              \frac{v^{\prime\prime}_3}{v^{\prime\prime}_1}
              \frac{v^{\prime\prime}_{\phi_2} }{ v^{\prime\prime}_{\phi_1} }  \\
               \frac{v^{\prime\prime}_{\phi_3} }{ v^{\prime\prime}_{\phi_1} } &
               \frac{v^{\prime\prime}_3}{v^{\prime\prime}_1}
               \frac{v^{\prime\prime}_{\phi_2} }{ v^{\prime\prime}_{\phi_1} }
               & \delta_3 \\
            \end{array}
          \right)\frac{f_\nu v^{\prime\prime}_{\phi_1} }{ \Lambda } \,v^{\prime\prime}_1,
          \label{neutrinos}
\end{equation}
in which $\delta_i$ is given by
\begin{eqnarray}
 \delta_1=\frac{f^\prime_\nu}{f_\nu}\,\frac{v_{_{T3}}v_{\phi_3}v_\chi}{\Lambda v_1v_{\phi_1}},\;
\delta_2=\frac{f^\prime_\nu}{f_\nu} \,\frac{v_{_{T1}} v_\chi}{ \Lambda v_1}, \; \delta_3= \frac{f^\prime_\nu}{f_\nu}
\,\frac{v_{_{T2}} v_{\phi_2} v_\chi}{ \Lambda v_1v_{\phi_1}},
\label{deltas}
\end{eqnarray}
where $\langle \chi\rangle=v_\chi$, e $\langle\Delta^0_i\rangle=v_{_{Ti}}$.

Since the mass matrix for the charged leptons is, for practical purposes,  diagonal, which
implies $ U^l_ {L, R} = 1 $, the mixing matrix PMNS, is defined as $V_{PMNS} =U^\nu \equiv U$, i.e., is obtained directly from
the diagonalization of the neutrino mass matrix.

In the neutrino sector the $V_{PMNS}$ matrix compatible with experimental data is, at the $3 \sigma$ level~\cite{concha},
\begin{equation}
| U |_{ 3 \sigma} = \left(\begin{array}{ccc} 0.77 - 0.86 & 0.50 - 0.63 & 0.00 - 0.22 \\ 0.22 - 0.56 & 0.44 - 0.73 &
0.57 - 0.80 \\0.21 - 0.55 & 0.40 - 0.71 & 0.59 - 0.82 \end{array}\right),
\label{u3s}
\end{equation}
We use a different strategy from that used with quarks. Imposing that the matrix $U$ satisfies the equation
$ U^T M_{\nu} U =~\hat{M}_{\nu}=~\textrm{diag}(m_1,m_2,m_3)$, where $M_\nu$ is given by the matrix in (\ref{neutrinos}),
we found numerical values for all the parameters in this matrix. Next, we found the square mass differences. For example,
using the lower limit of the entries in (\ref{u3s}), and the values of $\Lambda = 1$ TeV, $v_{\chi} = 2$ GeV and
$v_{T_1} = v_{T_2} = 2 v_{T_3} =1.25$ GeV and (in eV) $v_{1} = 4, v_{2} =10^{5}, v_{3} = 0.04, v_{\phi_1} = 40450,
v_{\phi_2} = 48000, v_{\phi_3} = 34850$, we obtain (in eV)
\begin{equation}
\hat{M}_\nu \approx \left(\begin{array}{ccc} 0.03485 &  \sim 0 &
    0 \\  \sim 0 & 0.0337083 & \sim 0 \\ 0 & \sim 0 & 0.06 \end{array}\right),
    \label{nus}
\end{equation}
where $\sim0$ means entries smaller than $10^{-7}$ eV.
From (\ref{nus}), the mass squared differences (in $\textrm{eV}^2$) obtained are
\begin{eqnarray}
\Delta m^2_{21}= 8 \times 10^{-5} ,\;\;
\Delta m^2_{31}= 2.5 \times 10^{-3},
\end{eqnarray}
which agree within 1$\sigma$ with the values at the $\mu=Z$ given in Ref.~\cite{massas}. The neutrino sector in the model differs from that
of the private Higgs of \cite{porto2} but, both models can accommodate a nonzero $\theta_{13}$ angle.
Notice that if we assume that $L(H^{\prime\prime})=L(\mathcal{T})=-2$,
and the other scalars having $L=0$, the interactions in (\ref{lagrangianaleptons}) conserve the lepton number.
However, there is no Majoron since the $A_4$ and $Z_3$ symmetries allow terms in the potential which break
explicitly the lepton number. For instance,  $[H^{\prime\prime \dag}\Phi]_1[\hat{H}^{\dag} H]_1$,
$[\hat{H} \epsilon \Phi^\prime]_1[H^{\prime} \epsilon H^{\prime\prime}]_1$, $[H^{ \dag} H^{\prime\prime}]_1
[\hat{H} \epsilon H^{\prime}]_1$. As in the quark sector, there are also FCNCs in the lepton sector but they will be considered elsewhere.

\textsl{Conclusion}. Motivated by the different mass scales in the quark and lepton sector we propose a model in which
each charge sector has its own Higgs scalars which acquires  appropriate VEVs. In this case, in order to induce the respective fermion
masses, the Yukawa couplings do not need to have a large hierarchy among them, i.e., all of them may be of the same
order of magnitude~\cite{dim5}. The hierarchy is translated to the values of the VEVs but these could be, in principle,  explained by the
minimization of the scalar potential. In fact, it was shown in \cite{3Dmodel} that this is indeed the case, for three scalar doublets.

Matrices like those in (\ref{massmatrices1}) are written in terms of
dimensionless parameters. This is because in the context of the SM, where these \textit{ansatze} have been
usually studied, the mass scale is already determined by the VEV $v_{_{SM}}=(G_F/2\sqrt2)^{1/2}$ but in multi-Higgs
models only $\sqrt{\sum_iv^2_i}$ which satisfies this constrained. This is the case of the model
in Ref.~\cite{dim5} that we are considering here, and all VEVs are considered parameters to be fixed only by
the fermion masses in each charge sector. We recall that we are not considering physical phases, but
$V^{U,D}_{L,R}$ may have the number of phases that are allowed for an arbitrary unitary matrix. We allow that in the
$V_{_{CKM}}$ matrix only one physical  $CP$ violating phase to survive after redefining the quark fields.

Another concern is if the suppression of the FCNC effect at low energies is stable up to higher energies.
The answer is yes, at least for the case of a general two Higgs doublet model~\cite{cvetic}. Moreover, QCD
corrections~\cite{gilman} have also to be considered at the next-to-leading order~\cite{next,next1}. This has been done
in the 2HDM in the context of natural flavor conservation and minimal flavour violation in Ref.~\cite{buras}.
A similar analysis in the context of models with at least three scalar doublets
with or without extra symmetries, say $A_4$, $S_3$, will be considered elsewhere.

\acknowledgments

The authors would like to thank for fully support to CNPq (ACBM) and for partial
support to CNPq and FAPESP (VP).

%One of the authors (ACBM) would like to thanks full support to CNPq. This works was also
%partially supported by CNPq under the contract and 302102/2008-6 (VP).

%%%%%%%%%%%%%%%%%%%%%%%%%%%%%%%%%%%%%%%%%%%%%%%%%%%%%%%%%%%%%%%%

\end{document}